 \documentstyle[preprint,prb,aps]{revtex}  % DON'T CHANGE
\begin{document}                % INITIALIZE - DONT CHANGE

\title{Magnetic Properties of the Transition to  Localized Superconductivity 
around Columnar Defects}

\author{  Mauro M. Doria$^a$, and Sarah C.B. de Andrade$^b$}
\address{ 
$^a$ Instituto de F\'{\i}sica, Universidade Federal do Rio de Janeiro,
C.P. 68528  Rio de Janeiro 21945-970 RJ, Brazil.\\
$^b$ Departamento de F\'{\i}sica, Pontif\'{\i}cia Universidade 
Cat\'olica do Rio de Janeiro, Rio de Janeiro 22452-970 RJ, Brazil.\\}
\maketitle
%\large

\begin{abstract}

We consider flux lines in presence of a dense square array of columnar defects, with insulating core and radius of the order of the coherence length.
The properties of the phase transition from the flux line regime to the localized superconducting state are determined here. We show that the localized superconducting state is a new vortex state because of the counterclockwise and of the clockwise supercurrent flows found around each defect. We propose here a general  operator to count the vorticity of any  cavity inside a superconductor and give a new method  to obtain the reversible magnetization of superconductors with  multiply connected geometries. 

\end{abstract}

\par PACS numbers:\quad 74.20.De \quad 74.25.Dw \quad 74.60.Ge
\newpage
% .......................................................  

% ............... introduction ................................

Recent improvements in the resolution of electron beam lithography have allowed the fabrication of  ordered arrays of {\it columnar defects} ({\bf CD}) with radius $R$ not much larger than the coherence length $\xi$\cite{BAERT,LIN,MARTIN}.
{\it Flux lines} ({\bf FL}) in presence of a regular  lattice of CD show new collective properties such as matching and multiple quantum trapping.
Matching occurs every time the artificial CD
and FL lattices are commensurate.
According to some theories\cite{MS}, CD can trap many FL\cite{BUZDIN} up to the saturation limit of $R/2\xi(T)$, where $T$ is the temperature.

A regular array of CD turns the superconductor into a  multiply connected geometry. 
This system presents the interesting property of surface superconductivity 
around the CD edges, which is crucially dependent on the properties of the medium exterior to the superconductor,
as found  long ago by Saint-James and DeGennes\cite{SJDG}.
This medium, played here by the CD core, must be insulating such 
that the supercurrent flow is constrained to the superconductor's interior\cite{MAURER}.
For  a non-perforated sample above the upper critical field $H_{c2}$ superconductivity disappears in the bulk. 
However in case of a perforated sample, such as the case of a regular array of CD,  superconductivity remains above $H_{c2}$ around the CD edges, in the form of a thin layer of thickness $\xi(T)$, and this lasts until the field $H_{c3}$ is reached.
Bezryadin, Buzdin and Pannetier\cite{BBP}(BBP) determined several properties of this {\it localized superconducting state} ({\bf LSS}), such as the ratio $H_{c3}/H_{c2}$ whose large $R$ limit recovers the single planar interface problem, first considered by Saint-James and DeGennes.
BBP also found that there are many possible LSS, each  characterized by $P$, the number of FL trapped inside the CD.
They determined that the transition between these LSS yields discontinuities in the magnetization (first order transition).
Since they assume the LSS around each CD from the beginning, their conclusions only apply for fields above the transition.
The interaction between the FL lattice and a single CD
was also investigated below the transition  where 
the  local symmetry group caused by  the deformation was determined\cite{BRAVERMAN}.
% ................ proposal ............................

In this letter we obtain the magnetic properties of the transition to the LSS, and
propose a Ginzburg-Landau free energy expansion that describes
the interaction between  FL and a dense regular lattice of CD parallel to them with radius size typically of the order of the  coherence length. 
The method also describes FL trapping by the CD,
which shows a  saturation maximum.
In particular we determine the magnetic properties (Fig[\ref{fig1},\ref{fig2}]) of the FL matter in both sides of the transition, which 
is signaled in the magnetization by  a slope discontinuity (second order transition).
We propose a quantity, hereafter called $f_{inh}$ (Eq.(\ref{finh})), 
to count the number of FL trapped inside a CD for any  $B$ and $T$.
Fig[\ref{fig2}] shows that $f_{inh}$ also determines the saturation number of FL inside the CD.
$f_{inh}$ is  a spatial average value over the order parameter and of its gradient, that picks its non-vanishing contribution  around the CD edge.
In Fig[\ref{fig3}] we show the strip  where  the LSS is  found in the diagram of the magnetic induction $B$ vs. $T$.

All results are obtained from numerical solutions of the present Ginzburg-Landau free energy expansion using the Simulated Annealing method\cite{KIRKPATRICK}, which was first applied to superconductivity some time ago\cite{DGR2}.
The magnetic properties were calculated using  the Virial Theorem\cite{DGR1,DS}, whose original formulation already included the treatment of  spatial inhomogeneities\cite{DGR1}.
The present treatment is for a regular square lattice of CD of  side $L$, with a number $N$ of FL inside the cell.
FL  are straight, thus there is translational invariance along the z-axis, and the problem is truly  two-dimensional. 
The  magnetic induction  becomes $\vec B = (1/V) \int dv \;\vec \nabla \times \vec A$ $=$ $ \hat z \; N \Phi_0/L^2$.
We chose to work with a single CD ($R=2.0\xi_0$) per unit cell ($L = 20\xi_0$), located at its center.  
This choice of unit cell does not allow the study of matching fields, because there is always commensurability between the two lattice parameters.
We work in the approximation of no screening, rendering this study interesting  for an extremely high $\kappa$ conventional superconductor such that London's penetration depth is much larger than the distance between two consecutive CD.

% ............... state description .............................

The onset of the LSS is shown in Fig[\ref{fig4}],
which displays the local supercharge and the local kinetic energy  densities.
Here $B$ is fixed ($N=24$), and  two temperature are considered,
$T/T_c =$ $0.615$ and  $0.630$,
below and very near the transition, respectively.
The transition occurs at $T/T_c= 0.631$.
For both temperatures $|\psi|^2$  is highly concentrated in the vicinity $\xi(T)$ around the CD edge, showing the onset of the LSS.
For $T/T_c =0.615$ small depressions of $|\psi|^2$
can still be  identified outside the localized state around the CD.
These depressions do correspond to FL external to the CD, because they coincide with local phase singularities and have counterclockwise supercurrent flow around them; these last two properties are the true signatures for the presence of a FL.
In fact the high symmetry on the arrangement of external  FL, formed around the CD below the transition, is a demand of the Saint-James and DeGennes condition.
Clearly, symmetry of the external FL configuration around the CD implies that the total supercurrent  vanishes at the CD center\cite{BRAVERMAN}.
The supercurrent flow is best understood through the local kinetic energy density.
For $T/T_c =0.615$ there is an internal ring and several spikes around the CD,
forming a highly symmetrical arrangement.
For $T/T_c =0.630$ the internal ring remains, but the kinetic density spikes collapse into a second ring.
We have determined that this external kinetic ring is caused by a clockwise supercurrent flow, thus opposite to the flow of the inner original ring, whose counterclockwise flow  is due to the $P$ trapped FL inside the CD.
Notice that the two kinetic energy density rings are located at the inner and the outer sides of  $|\psi|^2$ in the LSS.

%..................... model properties ................................

We describe the superconductor with  a square lattice of CD through
the free energy density expansion $ f = F - {\vec B}^2/8\pi$ $=$ $ (1/V) \int dv \lbrace \; \hbar^2 | \vec D \psi|^2/2 m(\vec x) + \alpha_0 \lbrack T - T_c(\vec x)\rbrack |\psi|^2 + \beta |\psi|^4/2 \; \rbrace$, $ \vec D \equiv \vec \nabla - (2\pi i /\Phi_0) \vec A$, 
$T_c(\vec x) = \lbrack 1 - \tau(\vec x) \rbrack T_c$, and $1/m(\vec x) = \lbrack 1 - \tau(\vec x) \rbrack/m $.
The CD are described in this model through the $\tau(\vec x)$ function, $\vec x$ being the distance from a CD center: $\tau(\vec x) = 1$ inside and $\tau(\vec x) = 0$ outside the defect, respectively.
To implement  this numerical study, the following function is introduced,
\begin{eqnarray}
\tau(\vec x) = 2 / \big(\; 1 + \exp{{\lbrack(|\vec x|/R)}^K\rbrack}\;\big),
\label{tauf}
\end{eqnarray}
where  $K$ is a phenomenological parameter, such that the Saint-James and DeGennes\cite{SJDG}  condition corresponds to the limit $K \rightarrow \infty$. For the present numerical study we take $K=5$.

It is convenient to introduce reduced units:
the order parameter becomes $\Delta \equiv \sqrt{\beta/\alpha_0(T_c-T)}\psi$;
both magnetic induction $( B')$ and magnetization $( M')$ are in units of $\sqrt{2} H_c(T)$, ${H_c(T)}^2/4\pi \equiv \alpha_0^2(T_c-T)^2/\beta$. 
The average energy densities are in units of ${H_c}^2(T)/4\pi$:
$f'$, $f'_{kin}$, and $f'_{inh}$.
To eliminate  unnecessary temperature dependence, the following dimensionless  fields are introduced: $ B \equiv (1-T/T_c) B' = (2\pi N/L^2) \xi_0^2 \kappa$, $\kappa \equiv \Phi_0/2\pi \sqrt{2} {\xi(T)}^2 H_c(T)$, $ M \equiv (1-T/T_c) M'$, $ f_{inh} = (1-T/T_c)^2 f'_{inh}$, etc.
In reduced units the free energy becomes,
\begin{eqnarray}
f' &=&  f'_{kin} +{1\over V} \int dv \; \lbrace \; - \lbrack 1 - \tau(\vec x)/(1 - T/T_c) \; \rbrack
|\Delta|^2 + |\Delta|^4/2 \; \rbrace \label{freen}\\ f'_{kin} &=& {1\over V} \int dv \;
\lbrack \; 1 - \tau(\vec x) \; \rbrack {\xi(T)}^2 |\vec D \Delta |^2,
\label{fkin}
\end{eqnarray}
where $\xi(T) = \xi_0 /\sqrt{1-T/T_c}$, $\xi_0 = \sqrt{\hbar^2/2m\alpha_0 T_c}$.
The corresponding Euler-Lagrange equation is
\begin{eqnarray}
{\xi(T)}^2 \; (-\vec \nabla \tau ) \cdot \vec D \Delta = - (1-\tau) {\xi(T)}^2 {\vec D}^2 \Delta - \lbrack 1 - \tau  /(1-T/T_c)\rbrack \Delta + |\Delta|^2 \Delta.
\label{eula}
\end{eqnarray}
For $K \rightarrow \infty$, $\tau$ approaches the step function
and so, in this limit, $\vec \nabla \tau \rightarrow -(R/|\vec x|) \delta(|\vec x|-R)
\hat r$. 
A physical solution of this equation only exists if its left side is finite and this demands that $\hat r \cdot \vec D \Delta|_{|\vec x| = R} = 0$, which is the Saint-James  and DeGennes boundary condition.
This shows that the present theory provides a method to implement this boundary condition  at the free energy level (Eq.(\ref{freen})).
The  Virial Theorem, known to provide a simple method to understand Abrikosov's theory\cite{KP},  gives the magnetization in this approximation of no screening, 
\begin{eqnarray}
4 \pi M' &=& \big( - f'_{kin}/2 + f'_{inh}/4 \big)/B'  \label{mag}\\
f'_{inh} &=& {1\over V} \int dv \big(-\vec x \cdot \vec \nabla \tau \big) \lbrack
{\xi(T)}^2 |\vec D \Delta |^2 - |\Delta|^2 / (1 - T/T_c) \rbrack, \label{finh}
\end{eqnarray}
where $\Delta$ is  a solution of Eq.(\ref{eula}).
$f'_{inh}$, the counter of trapped FL, directly contributes to the magnetization, thus  explaining why the transition is of first order when the number of  trapped FL changes inside the CD.
The step function limit  clearly shows that the major contribution to $f'_{inh}$ (Eq.(\ref{finh})) comes just from
the supercharge and gradient densities close to the  CD edge.

%....................numerical solution................

To  find the solution of  Eq.(\ref{eula}), first Eq.(\ref{freen}) is discretized, and  then its minimum numerically sought through the Simulated Annealing method\cite{DGR2}. 
There are ${J}$ values of the order parameter to be determined: $\Delta(n_x,n_y)$,
$n_x = 1,\ldots, J$, and $n_y=1,\ldots, J$, where
the new  working parameter, $a=L/J$\cite{NOTA1}, is the distance between two consecutive points.
All the present numerical results were obtained using a $J=30$ array of points, so that $a= (2/3)\xi_0$\cite{NOTA2}.

%................figure discussion ..........................

Our major results are displayed in the Figures.
Fig[\ref{fig1}] shows $M$ vs. $B$ for $T=0$, corresponding to   $N=1, \dots, 85$
FL in the unit cell, and the inset shows the transition region.
At the transition to the LSS ($N=68$) an abrupt change of slope in the magnetization takes place. 
The discontinuities in the magnetization, signaled by 
$B_P$, $P=2,3,4, 5$, are the transitions to new trapped states, characterized by $P$ FL inside.
The $B_5$ transition falls above the LSS transition, 
showing that multiple FL trapping and the LSS
are two independent transitions.
In fact the $B_5$ transition has been utterly discussed by BBP\cite{BBP}.
The $B_P$ transitions are best understood through 
Fig[\ref{fig2}], which shows the diagram $f_{inh}$ vs. $B$.
Notice that this curve naturally splits into five distinct lines, each associated to a $P$ trapped state.
For instance, for $N=1, \ldots, 5$ FL in the unit cell, and also for $N=9$
one has that  $P=1$. But for $N=6,7,8$,  there are $P=2$ trapped FL inside the CD. 
We believe that re-entrant effects are caused by the numerical procedure, which cannot  resolve the very fine
splitting between two possible states very close in energy, but distinct in 
the number of trapped FL.
This difficulty is well resolved for  $B_3$ and $B_5$,
but not for $B_2$ and $B_4$.
The inset of Fig[\ref{fig2}] shows $f_{inh}$ vs. $B$ for a higher temperature
($T/T_c=0.6$), and is quite similar to the $T =0$ plot, except for the fact that the destruction of superconductivity happens at a lower field ($B/\kappa \approx 0.5$), which limits the highest trapped state to $P=3$.
Fig[\ref{fig3}] shows that the LSS is found between the two data sets, which
were obtained by keeping $T$ fixed and varying $N$.
Above $T/T_c=0.8$ there is no more LSS since  CD interact collectively. 
The inset of Fig[\ref{fig3}] shows the same discontinuity on the magnetization slope as the inset of Fig[\ref{fig1}], but now obtained by varying $T$ 
and keeping  $N$ fixed instead.

%..................... zero current ...................................

Very near to a CD edge, the  supercurrent flow  is counterclockwise,
both below the transition and also in the LSS.
Remarkably, in the LSS,  away from the CD edge, the supercurrent flow flips direction and becomes clockwise. Thus it vanishes at a critical radius $|\vec x_0|$  from the CD center.
Our numerical data is in agreement with a simple theoretical argument, given below, that this critical radius is
\begin{eqnarray}
|\vec x_0| = \sqrt{P/\pi N}\; L. \label{x0}
\end{eqnarray}
Below the transition consider  the distance from the CD center to the symmetrical FL arrangement around the CD.
External FL are attracted to the  CD center, because CD are pinning centers, and also repelled from them, because of the multi FL  already trapped there, resulting on an equilibrium distance.
This equilibrium occasionally can change when a new FL enters  the CD interior.
This process does not often occur because of the saturation limit, which is low for a CD with  $R \sim \xi_0$.
Once the absorption process is completed, a new equilibrium symmetrical configuration of external FL is established.
We find that when the transition line to the LSS is approached this equilibrium distance grows, meaning that near the CD edge the repulsion caused by the trapped FL becomes stronger and the pinning attraction to the CD weaker.
The   collective clockwise supercurrent flow arises between  this symmetrical FL configuration  and the CD edge.
Within the LSS, and under the assumption that
all localized states are independent,
the wave function around each  CD is approximately given by $\psi = |\psi(|\vec x|)|\exp{(iP\theta)}$, where $|\vec x|$ is the distance to the CD center,
and $P$ FL are trapped inside.
Under the assumption of no screening in the unit cell $(A_x=0, A_y=B x)$,
the supercurrent associate to this state, $\vec J \propto |\psi(|\vec x|)|^2 (P - \pi |\vec x|^2 B/\Phi_0) \hat \theta$, flips sign at a distance given by Eq.(\ref{x0}). 
Since $|\vec x_0|$ coincides with  the peak of $|\psi|^2$, it must be of the order of $\xi(T)$ away from the CD edge.

%...................summary......................

In summary we have studied here 
a dense  FL lattice in presence of a dense square lattice of CD
in the limit of no screening.
We have obtained the magnetic properties of the transition to
surface superconductivity around the CD.
The interaction between the saturated CD and the external FL,
the eventual   capture of a FL by the CD are all described here
through a Ginzburg-Landau free energy, numerically solved by the Simulated
Annealing method.

%.............acknowledgments.....................
%\section*{Acknowledgments}
%\footnotesize{We thank W. Morgado, G. Carneiro, S. Segui, and L. Ghivelder for helpful discussions, and one of us (M.M.D.) also thanks  CNPq and FAPERJ (Brasil) for financial support.}

%...............................................

\newpage
	     
\baselineskip = 2\baselineskip  % double space the text
\begin{figure}
% fizeau1.eps
%\setlength{\unitlength}{1mm}
%\begin{picture}(200,190)
%\put(0,50){\epsfxsize=15cm\epsfbox{fa1.eps}}
%\end{picture}
 \caption{The $M$ vs. $B$ curve is shown  for the parameters $R=2\xi_0$,
$J=30$, $L=20\xi_0$, $T =0$, and $K=5$. 
The number of FL in the unit cell varies from $N=1$ to $N=85$.
The  fields $B_P$ mark the transition to $P$  trapped FL inside the CD.
The inset shows the transition at $N=68$.}
\label{fig1}
\end{figure}
%------------------------
\begin{figure}
% fizeau2.eps
%\setlength{\unitlength}{1mm}
%\begin{picture}(200,190)
%\put(0,50){\epsfxsize=15cm\epsfbox{fa1.eps}}
%\end{picture}
 \caption{The $F_{inh}$ vs. $B$ curve splits into five independent lines
each one associated to $P$ trapped FL inside the CD.
The same parameters of Fig.[1] are used here.
The inset shows the same quantity for the higher temperature of $T/T_c=0.6$.}
\label{fig2}
\end{figure}
%------------------------
\begin{figure}
% fdelb.ps
\caption{The two data sets correspond to the onset of localized superconductivity (white circle) and  its disappearance (black circle).
The inset shows the transition for $B=0.377$, which occurs
for $T/T_c = 0.631$.}
\label{fig3}
\end{figure}
%------------------------
\begin{figure}
% fdelb.ps
\caption{ The local supercharge ($n_s= |\Delta|^2 $) and the local kinetic energy ($n_k=\lbrack \; 1 - \tau(\vec x) \; \rbrack {\xi(T)}^2 |\vec D \Delta |^2$) densities are shown here for the same $B$ ($N=24$), but for two different temperature  $t=T/T_c$, $t=0.615$ and $t=0.630$, below and very close the transition to the localized superconducting state.}
\label{fig4}
\end{figure}
%------------------------

\begin{thebibliography}{9}
\bibitem{NOTA1} In the discrete space integration becomes $\int (dv/V) \rightarrow (1/N_{p}^2)\sum_{n_{x}=1}^{N_{p}} \sum_{n_{y}=1}^{N_{p}}$. Discretized gauge invariant derivatives are given in these reduced units by $D_y \Delta \rightarrow (\Delta'- \exp{(2\pi i A_y a/\Phi_0)}\Delta)/a$, and $D_x\Delta \rightarrow (\Delta''-\Delta)/a$. 
Our choice of gauge is $(A_x=0, A_y =(N\Phi_0/J L)\;n_x $).
In  the kinetic energy the functions $(\tau+\tau')/2$ and $(\tau+\tau'')/2$ are used, instead of just $\tau$, so to keep the symmetry with respect to nearest neighbor points. 
The functions ($\tau'$,$\Delta'$) and ($\tau''$,$\Delta''$) refer to 
the upper and right nearest neighbor sites of ($\tau$,$\Delta$).
\bibitem{NOTA2} The present Simulated Annealing calculation was done with a typical value of $1,600$ Monte Carlo visits to each site at a given temperature reduction.
\end{thebibliography}
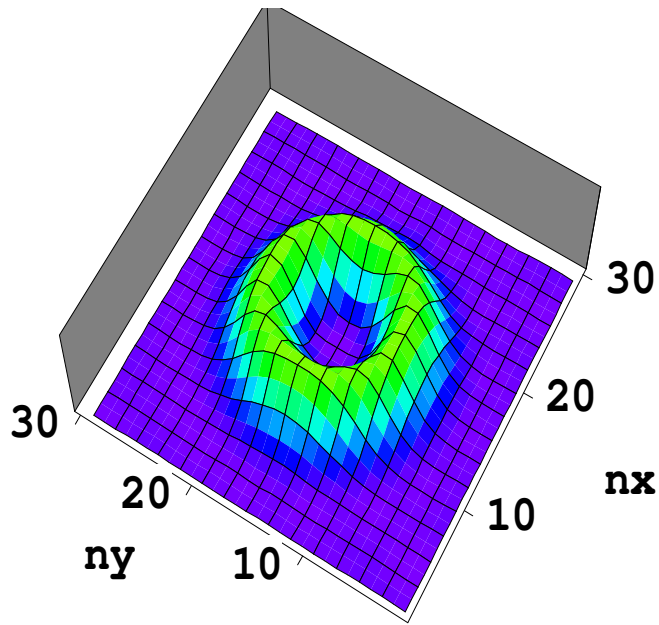
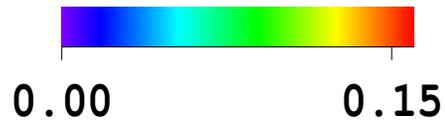
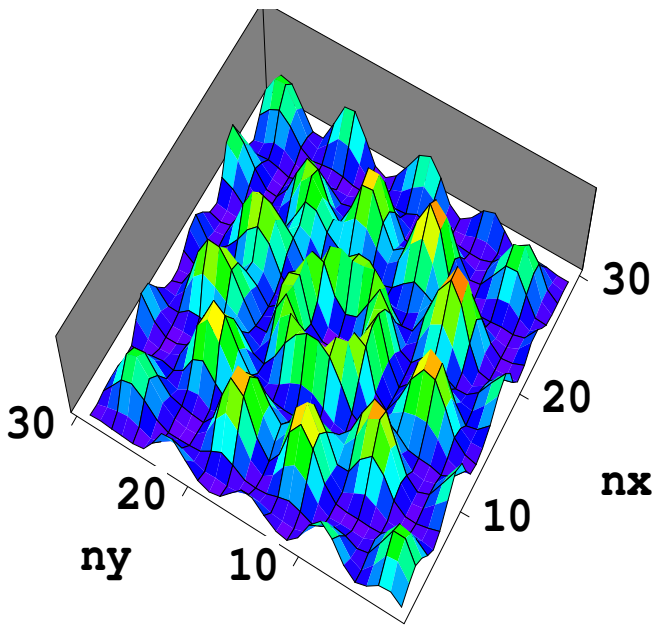
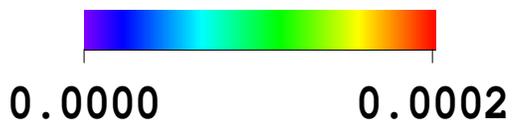
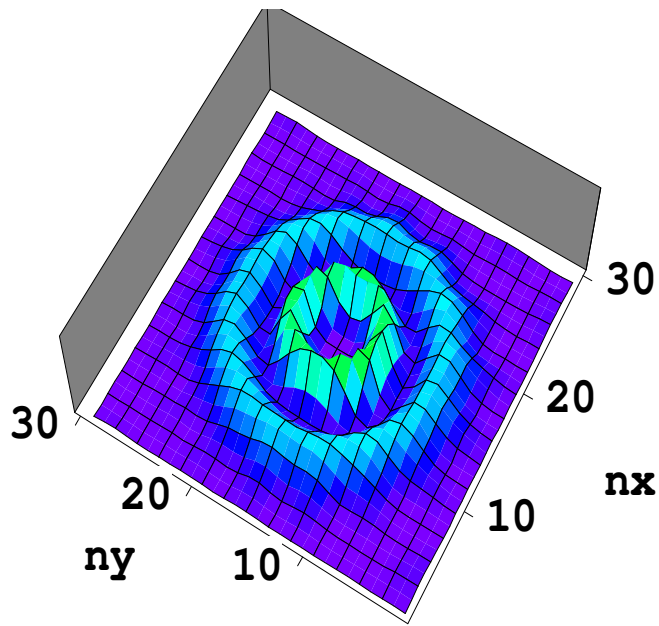
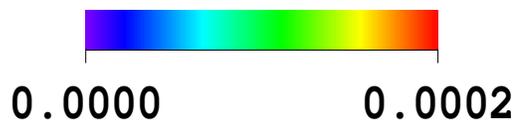
\end{document}